\documentclass{article}
\usepackage{amssymb,amsmath,graphicx,cite,bm}

\begin{document}

\title{Simulation of spherulite growth using a comprehensive approach to modeling the first-order isotropic/smectic-A mesophase transition}
\author{
        Nasser Mohieddin Abukhdeir and Alejandro D Rey\\
        Dept. of Chemical Engineering\\ 
          McGill University, Montreal, Quebec H3A 2B2\\
          Canada
}
\date{\today}

\maketitle

\begin{abstract}
A comprehensive modeling and simulation study of the first-order isotropic/smectic-A transition is presented and applied to phase diagram computation and two-dimensional spherulite growth.  An approach based on nonlinear optimization, that incorporates experimental data (from 12CB, dodecyl-cyanobiphenyl), is used to determine physically realistic model parameters.  These parameters are then used in conjunction with an optimized phase diagram computation method.  Additionally, a time-dependent formulation is presented and applied to the study of two-dimensional smectic-A spherulite growth.  These results show the growth kinematics and defect dynamics of nanoscale smectic-A spherulite growth in an isotropic phase with an initially radial layer configuration.
\end{abstract}

\section{Introduction} \label{sec_intro}
Liquid crystallinity and other forms of self-organization are key phenomena both technologically and in Nature.  Liquid crystalline order ranges from liquids that show some degree of orientational order to those that, in addition, show various types of translational order.  The myriad of types of material that exhibit this behavior range from traditional low molecular mass molecules, currently used in display technology, to biological membranes composed of phospholipids \cite{Fisch2004}.  To date, the main focus of liquid crystal research has been on the simplest of the class  of mesophases where some degree of orientational order is present: the nematics.  Research on mesophases that also show a degree of translational order, including smectics and columnar liquid crystals, has been less abundant.  Recognizing the increasing importance of these mesophases, particularly in biological systems, the need for practical methods to access the time and length scales at which these phenomena occur becomes more important.

Experimental work in this general field has made great progress in the basic understanding of translationally ordered liquid crystals \cite{Harrison2000,Harrison2002,Ruiz2007,Michel2006,Kleman2008}.  Nonetheless, it is currently infeasible to access much of the dynamic phenomena of translational phase-ordering processes.  Recent experimental work has begun to address these issues \cite{Ruiz2007,Michel2006}, but theoretical approaches are currently the only way to access the length scales (nanometers) and time scales (nanoseconds) at which liquid crystal dynamics occur.  The use of high-order models in conjunction with advanced numerical simulation techniques has shown a great deal of promise for theoretical study \cite{Mukherjee2001,Biscari2007,Abukhdeir2008,Abukhdeir2008a}.  Recent computational advances have allowed for the possibility of simulation in greater detail than ever before.

Utilizing a high-order Landau-de Gennes type model of the first-order isotropic/smectic-A mesophase transition \cite{deGennes1995,Mukherjee2001}, the objectives of this work are:
\begin{itemize}
\item to present a comprehensive approach to modeling and simulation of the first-order isotropic/smectic-A transition.
\item to determine phenomenological model parameters through incorporation of experimental data (from 12CB, dodecyl-cyanobiphenyl).
\item to efficiently compute a phase diagram predicted by the model and parameter set.
\item to study the two-dimensional growth kinematics and defect dynamics of an initially radial smectic-A spherulite in an isotropic matrix.
\end{itemize}
This approach builds upon previous work \cite{Abukhdeir2007}, which incorporates experimental data into the phenomenological model and derives equations for phase diagram computation.  A time-dependent formulation \cite{Abukhdeir2008,Abukhdeir2008a} and the nano-scale growth of an initially radial spherulite are presented.  This work is organized as follows: a brief background on relevant types of liquid crystalline order is given (Section \ref{subsec_lcorder}), the model and simulation approach are explained (Sections \ref{subsec_params}-\ref{sec_simcond}), and simulation results are presented and discussed (Section \ref{sec_resdisc}).

\section{Background and theory} \label{sec_back}

\subsection{Liquid crystalline order} \label{subsec_lcorder}

Liquid crystalline phases or mesophases are materials which exhibit partial orientational and/or translational order.  They are composed of anisotropic molecules  which can be disc-like (discotic) or rod-like (calamitic) in shape.  Thermotropic liquid crystals are typically pure-component compounds that exhibit mesophase ordering most greatly in response to temperature changes.  Lyotropic liquid crystals are mixtures of mesogens (molecules which exhibit some form of liquid crystallinity), possibly with a solvent, that most greatly exhibit mesophase behavior in response to concentration changes.  Effects of pressure and external fields also influence mesophase behavior.  This work focuses the study of calamitic thermotropic liquid crystals which exhibit a first-order mesophase transition.

\begin{figure}[htp]
\begin{center}
\includegraphics[width=4in]{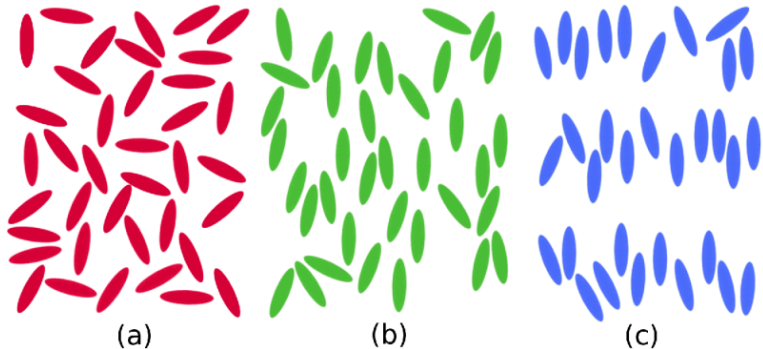} 
\end{center}
\caption{schematics of a) the isotropic, b) nematic, and c) smectic-A mesophases. \label{figlcorder}}
\end{figure}

An unordered liquid, where there is neither orientational nor translational order (apart from an average intermolecular separation distance) of the molecules, is referred to as isotropic.  Liquid crystalline order involves partial orientational order (nematics) and, additionally, partial translational order (smectics and columnar mesophases).  The simplest of the smectics is the smectic-A mesophase, which exhibits one-dimensional translational order in the direction of the preferred molecular orientational axis.  It can be thought of as layers of two-dimensional fluids stacked upon each other.  Schematic representation of these different types of ordering are shown in Figure \ref{figlcorder}.

\subsection{Order parameters and the phenomenological model} \label{subsec_elemdisl}

Theoretical characterization of mesophase order is accomplished using order parameters that adequately capture the physics involved.  These order parameters typically have an amplitude and phase associated with them.  In order to characterize the partial orientational order of the nematic phase, a second order symmetric traceless tensor can be used \cite{deGennes1995}:
\begin{equation} \label{eqnem_order_param}
\bm{Q} = S \left(\bm{nn} - \frac{1}{3} \bm{I}\right) + \frac{1}{3} P \left( \bm{mm} - \bm{ll}\right)
\end{equation}
where $\mathbf{n}/\mathbf{m}/\mathbf{l}$ are the eigenvectors of Q-tensor, which characterize the average molecular orientational axes, and $S/P$ are scalars which represent the extent to which the molecules conform to the average orientational axes \cite{Rey2002,Yan2002,Rey2007}.  Uniaxial order is characterized by $S$ and $\bm{n}$, which correspond to the maximum eigenvalue(and its corresponding eigenvector) of the Q-tensor, $S= \frac{3}{2} \mu_n$.  Biaxial order is characterized by $P$ and $\bm{m}/\bm{l}$, which correspond to the lesser eigenvalues and eigenvectors, $P = -\frac{3}{2}\left(\mu_m - \mu_l\right)$.

The smectic-A mesophase has one-dimensional translational order in addition to the orientational order found in nematics.  Characterizing this mesophase can be accomplished through the use of primary (orientational) and secondary (translational) order parameters together \cite{Toledano1987}.  This is accomplished using the tensor order parameter (\ref{eqnem_order_param}) and the complex order parameter \cite{deGennes1995}:
\begin{equation} \label{eqsmec_order_param}
\Psi = \psi e^{i \phi}
\end{equation}
where $\phi$ is the phase, $\psi$ is the scalar amplitude of the density modulation.  The density wave vector, which describes the average orientation of the smectic-A density modulation, is defined as $\mathbf{a} = \nabla \phi / |{\nabla \phi}|$.  The smectic scalar order parameter $\psi$ characterizes the magnitude of the density modulation, and is used in a dimensionless form in this work.  In the smectic-A mesophase the preferred orientation of the wave vector is parallel to the average molecular orientational axis, $\mathbf{n}$.

A Landau-de Gennes type model for the first order isotropic/smectic-A phase transition is used that was initially presented by Mukherjee, Pleiner, and Brand \cite{deGennes1995,Mukherjee2001} and later extended by adding nematic elastic terms \cite{Brand2001,Mukherjee2002a,Biscari2007}:
\begin{eqnarray} \label{eq:free_energy_heterogeneous}
f - f_0 &=&\frac{1}{2} a \left(\bm{Q} : \bm{Q}\right) - \frac{1}{3} b \left(\bm{Q}\cdot\bm{Q}\right) : \bm{Q} + \frac{1}{4} c \left(\bm{Q} : \bm{Q}\right)^2 + \frac{1}{2} \alpha \left|\Psi\right|^2 + \frac{1}{4} \beta \left|\Psi\right|^4 \nonumber\\
&&- \frac{1}{2} \delta \psi^2 \left(\bm{Q} : \bm{Q}\right) - \frac{1}{2} e \bm{Q}:\left(\bm{\nabla} \Psi\right)\left(\bm{\nabla} \Psi^*\right) \nonumber\\
&& + \frac{1}{2} l_1 (\bm{\nabla} \bm{Q} \vdots \bm{\nabla} \bm{Q} ) + \frac{1}{2} b_1 \left|\bm{\nabla} \Psi\right|^2 + \frac{1}{4} b_2 \left|\nabla^2 \Psi\right|^2
\end{eqnarray}

\begin{eqnarray} \label{eq:free_energy_heterogenous_coeffs}
A & = & a_0 (T - T_{NI}) \nonumber \\
\alpha & = & \alpha_0 (T - T_{AI})\nonumber 
\end{eqnarray}
where $f$ is the free energy density, $f_0$ is the free energy density of the isotropic phase, terms 1-5 are the bulk contributions to the free energy, terms 6-7 are couplings of nematic and smectic order; both the bulk order and coupling of the nematic director and smectic density-wave vector, respectively.  Terms 8-10 are the nematic and smectic elastic contributions to the free energy, respectively.  The order parameters are defined in (\ref{eqnem_order_param}-\ref{eqsmec_order_param}), $T$ is temperature, $T_{NI}$/$T_{AI}$ are the hypothetical second order transition temperatures for isotropic/nematic and isotropic/smectic-A mesophase transitions (refer to \cite{Coles1979a} for more detail), and the remaining constants are phenomenological parameters.  Further explanation and justification for the use of this high-order model can be found in \cite{Abukhdeir2008a}.

\subsection{Homogeneous free energy and parameter determination} \label{subsec_params}

Following past work \cite{Abukhdeir2007}, in order to compute the phase diagram from the free energy equation (\ref{eq:free_energy_heterogeneous}) a homogeneous uniaxially-ordered volume assumption can be used, resulting in a simplified point-volume free energy density:
\begin{equation} \label{eq:free_energy_homogeneous}
f - f_0  =  \frac{1}{3} a S^2 - \frac{2}{27} b S^3 + \frac{1}{9} c S^4 + \frac{1}{2} \alpha \psi^2 + \frac{1}{4} \beta \psi^4 - \frac{1}{3} \delta S^2 \psi^2 +  \frac{1}{2} b_1 \psi^2 q^2 + \frac{1}{4} b_2 \psi^2 q^4 - \frac{1}{3} e S \psi^2 q^2
\end{equation}
where $q=\frac{2 \pi}{d_0}$ is the magnitude of the wave vector and $d_0$ is the equilibrium layer spacing.  Note that a different definition of the Q-tensor (\ref{eqnem_order_param}) is used in this work than in ref. \cite{Abukhdeir2007}.

A major complication of applying this model is the determination of a suitable set of model parameters.  In order to overcome this challenge, a nonlinear programming formulation is derived which then allows to the application of nonlinear solution methods.  Utilizing experimental data \cite{Coles1979a,Urban2005,DasGupta2003} and minimization criteria of the homogeneous free energy (\ref{eq:free_energy_homogeneous}) \cite{Mukherjee2001,Prostakov2002a}, a nonlinear optimization problem is formulated:
\begin{eqnarray} 
obj&=& min\left(q_{l}^2 - q_{b}^2\right)\nonumber\\
\frac{\partial f(T_b)}{\partial X_i} &=& 0\nonumber\\
\frac{\partial^2 f(T_b)}{\partial S^2} &>& 0 \nonumber\\
\frac{\partial f(T_l)}{\partial X_i} &=& 0
\end{eqnarray} 
where the objective function $obj$ minimizes the change in the layer spacing between the unknown bulk transition value, $q_{b}$, and the known value at some minimum valid temperature for the model, $q_l$.  Each constraint is evaluated at the corresponding temperatures $T_b$, the bulk transition temperature, and $T_l$, the minimum valid temperature for the model parameter set.

\subsection{Phase diagram determination} \label{sec_phasediag}

\begin{figure}[htp]
\begin{center}
\includegraphics[width=4in]{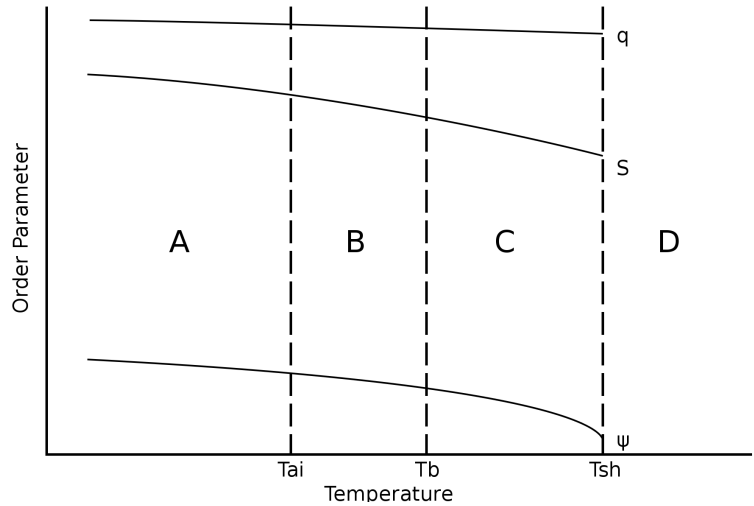} 
\end{center}
\caption{Phase diagram schematic of the first-order isotropic/smectic-A phase transition where $T_{sh}$ corresponds to the upper stability limit of a super-heated smectic domain, $T_b$ refers to the bulk transition temperature (the free energy of the isotropic and smectic-A phases are equal), and $T_{AI}$ refers to the lower stability limit of a super-cooled isotropic domain, also referred to as the theoretical second-order transition temperature \cite{Coles1979a}; the smectic-A phase is stable and the isotropic phase is unstable in region A, the smectic-A phase is stable and the isotropic phase is meta-stable in region B, the smectic-A phase is metastable and the isotropic phase stable in region C, and the smectic-A phase unstable and isotropic phase stable in region D. \label{figphasediagschem}}
\end{figure}

Utilizing the homogeneous free energy (\ref{eq:free_energy_homogeneous}) and model parameters determined from the nonlinear optimization solution in Section \ref{subsec_params}, a phase diagram for the system can be computed.  Figure \ref{figphasediagschem} shows a schematic of the phase diagram of a first-order isotropic/smectic-A phase transition.  Coexistence regions are present, due to the first-order nature of the transition, where both the isotropic and smectic-A phase are either stable or metastable. 

 Minimization of the homogeneous free energy (\ref{eq:free_energy_homogeneous}) is computationally challenging due to the presence of three degrees of freedom.  This can be simplified by parameterization of the free energy equation \cite{Mukherjee2001} using minimization invariants and the assumption that the smectic-A phase exists.  The resulting free energy equation, parameterized as a function of the nematic scalar order parameter (see (\ref{eqnem_order_param})), is:
\begin{eqnarray} \label{eqparamfreeen}
f_{A} & = & A S_A^4 + B S_A^3 + C S_A^2 + D S_A + E \nonumber \\
A & = & -\left(16 e^4+ 96 b_2 \delta e^2 + 144 b_2^2 \delta^2 - 144 b_2^2 \beta c \right) \left(1296 b_2^2 \beta \right)^{-1} \nonumber \\
B & = & -\left(-96 b_1 e^3- 288 b_1 b_2 \delta e+ 96 b b_2^2 \beta \right) \left(1296 b_2^2 \beta \right)^{-1} \nonumber \\
C & = & -\left( \left(216 b_1^2 - 144 \alpha b_2 \right) e^2 + \left(216 b_1^2 b_2- 432 \alpha b_2^2 \right) \delta - 432 a b_2^2 \beta \right)
\left(1296 b_2^2 \beta \right)^{-1} \nonumber \\
D & = & -\left( 432 \alpha b_1 b_2 - 216 b_1^3 \right) e \left(1296 b_2^2 \beta \right)^{-1}\nonumber \\
E & = & - \left(324 \alpha^2 b_2^2 - 324 \alpha b_1^2 b_2 + 81 b_1^4  \right) \left(1296 b_2^2 \beta \right)^{-1}
\end{eqnarray}
where $S_A$ is the nematic scalar order parameter with the assumption that the smectic-A phase is stable/meta-stable.  The minima of (\ref{eqparamfreeen}) are easily found in that they are the roots of a polynomial, $\frac{d f_{A}}{dS_A}$.  Once the minima are determined, the validity of the assumption that the smectic-A phase exists (at the specified temperature) can be tested by i) verifying that the computed $S_A$ and the corresponding values of $\psi$/$q$ are real, and ii) verification of the minimization criteria of the full homogeneous free energy (\ref{eq:free_energy_homogeneous}).  The equations to determine $q$ and $\psi$ from $S_A$ are derived from the application of the minimization criteria to the homogeneous free energy equation (\ref{eq:free_energy_homogeneous}):
\begin{eqnarray}
q^2 &=& \frac{2 e S_A-3 b_1}{3 b_2} \nonumber \\
\psi^2 &= & \frac{4 \delta S_A^2+4 e S_A q^2-3 b_2 q^4-6 b_1 q^2 - 6 \alpha}{6 \beta}
\end{eqnarray}

\subsection{Time-dependent formulation} \label{sec_timedep}

The Landau-Ginzburg time-dependent formulation\cite{Barbero2000} is used to capture the kinetics of the phase transition.  Due to the higher order derivative term in the free energy functional, a higher order functional derivative must be used.    Additionally, in order to utilize standard numerical solution techniques, the complex order parameter (\ref{eqsmec_order_param}) is separated into its real and imaginary contributions:
\begin{equation} \label{eqcomplex_order_split}
\Psi = A+Bi
\end{equation}
The general form of the time-dependent formulation is as follows \cite{Barbero2000}:
\begin{eqnarray} \label{eq:landau_ginz}
\left(\begin{array}{c}
 \frac{\partial \bm{Q}}{\partial t}
\\ \frac{\partial A}{\partial t}
\\ \frac{\partial B}{\partial t} 
\end{array}\right)
&=& 
\left(\begin{array}{c c c} 
\frac{1}{\mu_n} & 0 & 0\\ 
0 & \frac{1}{\mu_S} & 0\\ 
0 & 0 & \frac{1}{\mu_S}\end{array} \right)
\left(\begin{array}{c} -\frac{\delta F}{\delta \bm{Q}}\\ 
-\frac{\delta F}{\delta A}\\ 
-\frac{\delta F}{\delta B} \end{array}\right)\\
F &=& \int_V f dV
\end{eqnarray}
where $\mu_r$/$\mu_s$ is the rotational/smectic viscosity, $f$ is the heterogeneous free energy density (\ref{eq:free_energy_heterogeneous}), and $V$ the volume.  As previously mentioned, a higher order functional derivative must be used due to the second-derivative term in the free energy (\ref{eq:free_energy_heterogeneous}):
\begin{equation} \label{eqpdes}
\frac{\delta F}{\delta \theta} = \frac{\partial f}{\partial \theta} - \frac{\partial}{\partial x_i}\left(\frac{\partial f}{\partial \frac{\partial \theta}{\partial x_i}} \right) + \frac{\partial}{\partial x_i}\frac{\partial}{\partial x_j}\left(\frac{\partial f}{\partial \frac{\partial^2 \theta}{\partial x_i \partial x_j}} \right)
\end{equation}
where $\theta$ corresponds to the order parameter.

The use of the full Q-tensor in this time-dependent model does not neglect biaxiality as was assumed in Sections \ref{subsec_params}-\ref{sec_phasediag}.  Substituting (\ref{eqcomplex_order_split}), the free energy (\ref{eq:free_energy_heterogeneous}), and high order functional derivative (\ref{eqpdes}) into the time-dependent formulation (\ref{eq:landau_ginz}) yields the closed set of simulated equations:
\begin{eqnarray} \label{eqthemodel}
\frac{\partial \bm{Q}}{\partial t} &=& -\left[ a^* \bm{Q} - b^* \left( \bm{Q} \cdot \bm{Q} \right)^{ST} + c^* \left( \bm{Q}:\bm{Q}\right)\bm{Q} - \delta^* \left( A^2 +B^2\right) \bm{Q} - \frac{1}{2} e^* \left( \bm{\nabla}A\bm{\nabla}A + \bm{\nabla}B\bm{\nabla}B \right)^{ST}\right] \nonumber\\
&&+ \bm{\nabla} \cdot \left( l_1^* \bm{\nabla Q} \right) \nonumber\\
\mu^* \frac{\partial A}{\partial t} &=& -\left[ \alpha^* A + \beta^* \left( A^2 + B^2 \right)A - \delta^* A \left(\bm{Q}:\bm{Q}\right)\right] +\bm{\nabla} \cdot \left[  b_1^* \bm{\nabla}A - e^* \bm{Q} \cdot \bm{\nabla} A - \frac{1}{2} b_2^* \bm{\nabla} \left(\nabla^2 A\right) \right]\nonumber\\ 
\mu^* \frac{\partial B}{\partial t} &=& -\left[ \alpha^* B + \beta^* \left( A^2  B^2 \right)B - \delta^* B \left(\bm{Q}:\bm{Q}\right)\right] + \bm{\nabla} \cdot \left[  b_1^* \bm{\nabla}B - e^* \bm{Q} \cdot \bm{\nabla} B - \frac{1}{2} b_2^* \bm{\nabla} \left(\nabla^2 B\right) \right]\nonumber\\ 
&&
\end{eqnarray}
where the asterisk denotes an nondimensionalized value, the superscript $ST$ denotes the symmetric/traceless portion of a tensor, and $\mu^*$ is the ratio of the smectic and rotational viscosities.  The nondimensionalized model parameters are as follows:
\begin{align}
a^* & = \frac{a_0 \overset{-}{T}}{\alpha_0}& b^* &= \frac{b}{\alpha_0 \Delta T}& c^* & =\frac{c}{\alpha_0 \Delta T} &\alpha^* & = \overset{-}{T}-1 & \beta^* &= \frac{\beta}{\alpha_0 \Delta T}& \delta^* & =\frac{\delta}{\alpha_0 \Delta T}& \nonumber\\
b_1^* & = \frac{b_1}{l^2 \alpha_0 \Delta T}& b_2^* & = \frac{b_2}{l^4 \alpha_0 \Delta T}& e^* & = \frac{e}{l^2 \alpha_0 \Delta T}& l_1^* & = \frac{l_1}{l^2 \alpha_0 \Delta T}& \nonumber \\
\mu^* &= \frac{\mu_s}{\mu_n}& \tau & = \frac{\mu_n}{\alpha_0 \Delta T}& \overset{-}{T}& = \frac{T - T_{ni}}{\Delta T}& \Delta T &= T_{AI}-T_{NI}&
\end{align}
where $l$ is the simulation-specific imposed length scale.

\subsection{Simulation conditions and computational approach} \label{sec_simcond}

\begin{figure}[htp]
\begin{center}
\includegraphics[width=2.75in]{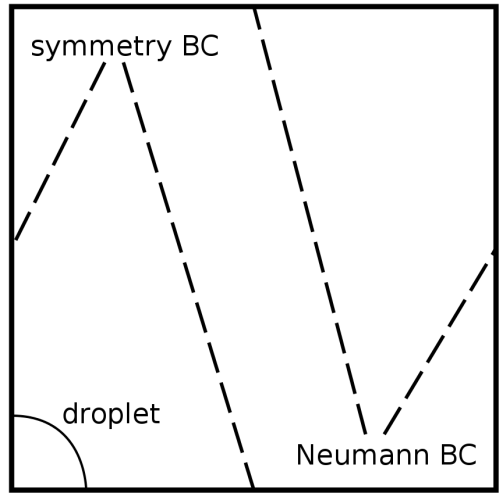} 
\includegraphics[width=2.75in]{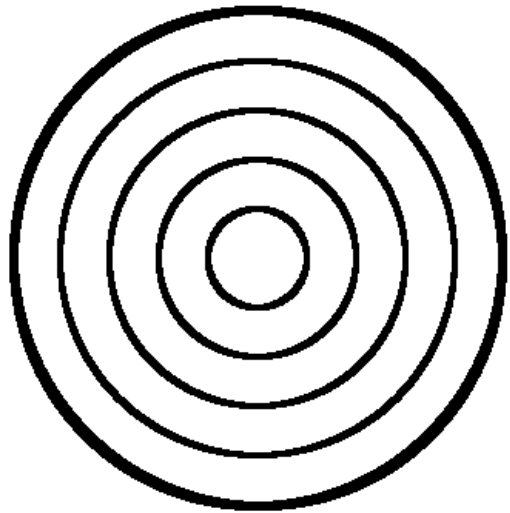} 
\end{center}
\caption{a) (left) a schematic of the simulation domain showing boundary conditions and spherulite placement b) (right) Schematic of the initially radial spherulite layer configuration used as an initial condition for simulation. The center of the radial spherulite is initially assumed isotropic. \label{figdroptexture}}
\end{figure}

A square computational domain with imposed length scales of $l=9.75\times 10^{-2} \mu m$ (approximately 25 layers) and $l=1.95\times 10^{-1} \mu m$ (approximately 50 layers) were used in two separate simulations.  Referring to Figure \ref{figdroptexture}a, both symmetry and Neumann boundary conditions were used to simulate bulk conditions.  Symmetry conditions for the Q-tensor (\ref{eqnem_order_param}) must  take into account vector symmetry, which results in the following boundary conditions \cite{Abukhdeir2008b}:
\begin{equation}
\frac{\partial Q_{xx}}{\partial x} = 0;\frac{\partial Q_{yy}}{\partial y} = 0;Q_{xy}=Q_{yx}=0
\end{equation}
The initial condition for both simulations was a smectic-A spherulite in an initially radial layer configuration (see Figure \ref{figdroptexture}b).  The radius of the spherulite was initially set to $r_0=12.0nm$, a value on the order of the layer spacing at $330K$ (Section \ref{sec_phasediag}), $d_0=3.9nm$ \cite{Urban2005}.  The initial value used for $S$, $\psi$, and the layer spacing correspond to the homogeneous values at $T=330K$, determined from the computed phase diagram.  The Heaviside step function was used to generate the initial spherulite.  The constraint that the spherulite does not impinge on the domain boundaries was verified post-simulation.

A commercial package, Comsol Multiphysics, was used to solve the time-dependent model (\ref{eqthemodel}).  Quadratic Lagrange basis functions were used for the Q-tensor variables and quartic Hermite basis functions used for the complex order parameter components.  Standard numerical techniques were utilized to ensure convergence and stability of the solution.  This platform does not support adaptive mesh refinement, thus a uniform mesh was used with a density of approximately $14.8$ nodes/$nm^2$ for the 25 layer simulation and $3.7$ nodes/$nm^2$ the 50 layer simulation.  Previous simulations using this model and numerical method have shown good agreement with both past experimental and theoretical findings \cite{Abukhdeir2008,Abukhdeir2008a}.  Additionally, exhaustive past work using this numerical method and the Landau-de Gennes model for the first-order isotropic/nematic phase transition \cite{Wincure2006,Wincure2007,Wincure2007a,Wincure2007b} has served to further validate this simulation approach.

\section{Results and discussion} \label{sec_resdisc}

\subsection{Phase diagram computation} \label{resdisc_phase}
\begin{figure}[htp]
\begin{center}
\includegraphics[width=4in]{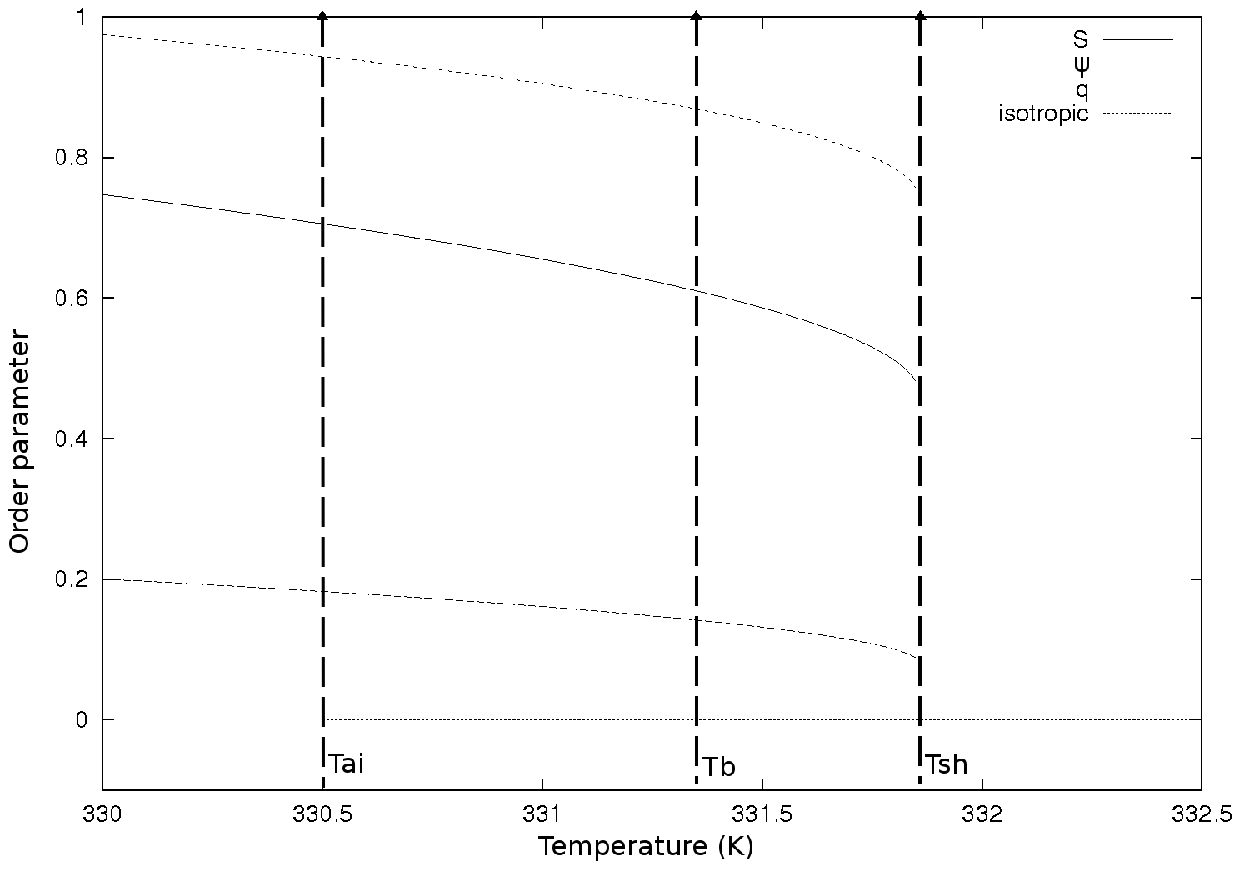} 
\end{center}
\caption{The computed phase diagram for the 12CB-based parameters.  Note that $q$, the wave vector magnitude (see (\ref{eq:free_energy_homogeneous})), is scaled by $\frac{2 \pi}{3.9\times 10^{-9}m}$ \cite{Urban2005}.  The experimental bulk transition temperature for 12CB is $331.35$ and the value of the nematic scalar order parameter at that temperature approximately $0.61$ \cite{Coles1979a}; The model parameters used in this computation and droplet growth simulations are as follows: $T_{NI}=322.85K$, $T_{AI}=330.5K$, $a_0= 2\times10^5\frac{J}{m^3 K}$, $b=2.823\times10^7\frac{J}{m^3}$, $c=1.972\times10^7\frac{J}{m^3}$, $\alpha_0=1.903\times10^6\frac{J}{m^3 K}$, $\beta=3.956\times10^8\frac{J}{m^3}$, $\delta=9.792\times10^6\frac{J}{m^3}$, $e=1.938\times10^{-11}pN$, $l_1=1\times10^{-12}pN$, $b_1=1\times10^{-12}pN$, $b_2=3.334\times10^{-30}Jm$, and the ratio of the rotational and diffusional viscosities used was $\frac{\mu_S}{\mu_N}=25$.  The rotational viscosity value $\mu_N = 8.4\times10^{-2}\frac{N \times s}{m^2}$ was used post-simulation for estimation of the time scale \cite{Wincure2006}. \label{figphasediag}}
\end{figure}

The algorithm for phase diagram computation presented in Section \ref{sec_phasediag} was implemented using a high-level numerical programming language \cite{Eaton2002}.  The resulting phase diagram is shown in Figure \ref{figphasediag}; refer to the phase diagram schematic, Figure \ref{figphasediagschem}, for a detailed explanation of the features.

The experimentally observed bulk transition temperature and nematic scalar order parameter \cite{Coles1979a} are exactly matched using the parameter determination method presented in Section \ref{subsec_params}.  Additionally, the smectic layer spacing on the order of $3.9nm$ \cite{Urban2005} is also well reproduced below the lower stability limit of the isotropic phase, $T_{ai}$.  Above this temperature the model predicts a layer spacing trend that increases approaching the super-heated stability limit of the smectic-A phase. 

\subsection{Two-dimensional spherulite growth}

\begin{figure}[htp]
\begin{center}
\includegraphics[width=6in]{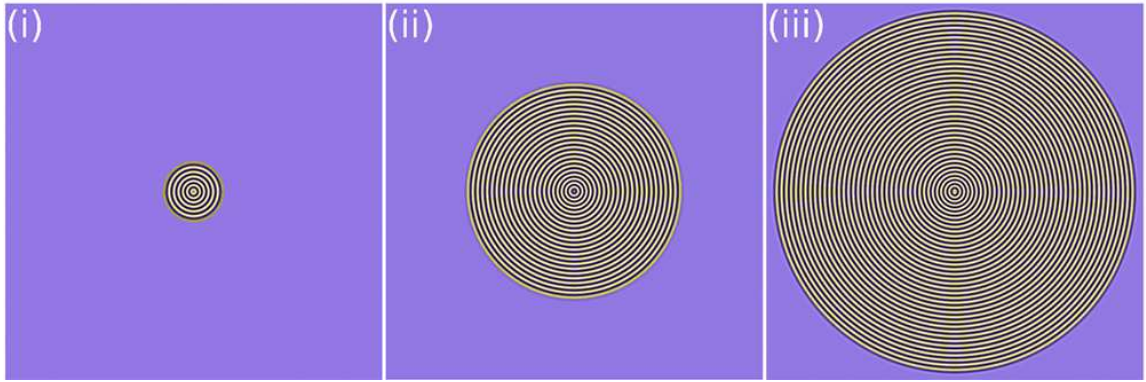} 
\end{center}
\caption{Results from the 50 layer simulation (domain length scale, utilizing symmetry, is $l=3.9\times 10^{-1} \mu m$) of the growth of an initially radial textured spherulite where the surface corresponds to $A = Re(\Psi)$; (i) $t = 3.40 \mu s$, the self-selected formation of a $+1$ disclination occurs from the initially isotropic core (see Section \ref{sec_simcond}) (ii) $t = 17.1 \mu s$, spherulite growth continues with the same core morphology (iii) $t = 28.7 \mu s$, growth continues with the splitting of the $+1$ disclination into two $+\frac{1}{2}$ smectic disclinations and the creation of a single disoriented smectic layer in the core; layer configuration results from the 25 layer simulation are indistinguishable from the 50 layer simulation, up to the maximum radius achieved with the 25 simulation, and are excluded for brevity. \label{figtext}}
\end{figure}

Simulation results from the growth of an initially radial spherulite are shown in Figure \ref{figtext} from the 50 layer simulation (see Section \ref{sec_simcond}).  Past work on the isotropic/nematic mesophase transition \cite{Yan2002,Sharma2003,Wincure2006,Wincure2007,Wincure2007a,Wincure2007b,Abukhdeir2008b} and smectic-A filament growth \cite{Rey2008a} provide a great deal of insight into the two general growth processes observed: shape dynamics \cite{Rey2007} and self-similar growth \cite{Sethna2007}.  As will be shown, these simulation results definitely show that the growth of an initially radial textured smectic-A spherulite in an isotropic phase follows a similar topological growth process as observed in its nematic counterpart \cite{Wincure2006,Wincure2007,Wincure2007a,Wincure2007b}.

\subsubsection{Shape dynamic growth}

\begin{figure}[htp]
\begin{center}
\includegraphics[width=6in]{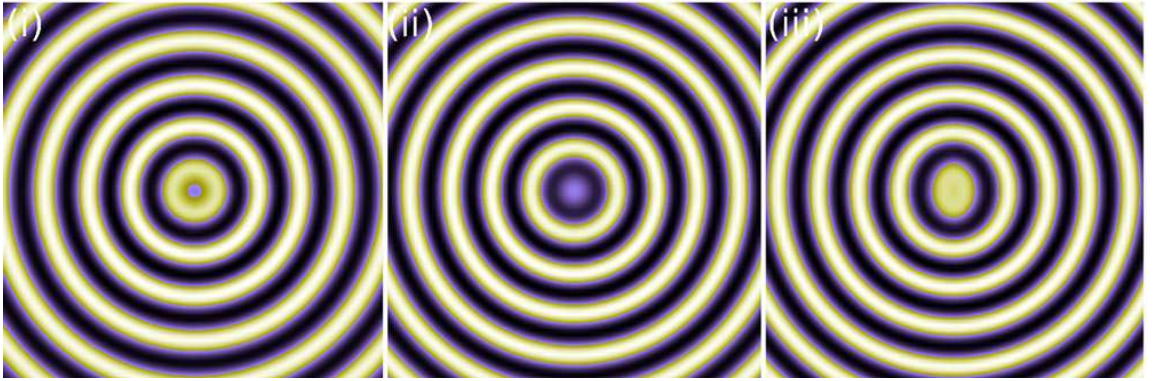} 
\end{center}
\caption{Enlarged results of the spherulite core from the 50 layer simulation (figure length scale is $l=39 nm$) corresponding to the full domain in Figure \ref{figtext} where the surface corresponds to $A = Re(\Psi)$; (i) $t = 3.40 \mu s$, the circular layer configuration in the vicinity of the $+1$ smectic disclination (ii) $t = 17.1 \mu s$, the splitting of the $+1$ disclination with the formation of a new layer (iii) $t = 28.7 \mu s$, the pair of $+\frac{1}{2}$ disclinations are fully formed which results in a single disoriented layer undergoing expansion with a wave vector orthogonal to the (vertical) axis formed by the two disclinations. \label{figtextzoom}}
\end{figure}

The shape dynamic growth regime is dominated by transient texturing and interfacial forces.  When the spherulite radius is on the order of the smectic coherence length:
\begin{equation}\label{eqn_smeclen}
 \lambda = \sqrt{\frac{b_1}{a_0 T}}
\end{equation}
long-range energy effects (gradients in molecular and layer orientation) are dominant over short-range energy effects (gradients in the bulk order parameters $S$ and $\psi$).  As the spherulite radius increases, interfacial anchoring affects the core of the spherulite less and short-range energy gradients become dominant in this region.  This results in increased texturing and defect dynamics \cite{Rey2002,Yan2002}.  Referring to Figure \ref{figtextzoom}, this process is observed for an initially radial smectic-A spherulite.

\begin{figure}[htp]
\begin{center}
\includegraphics[height=2.75in]{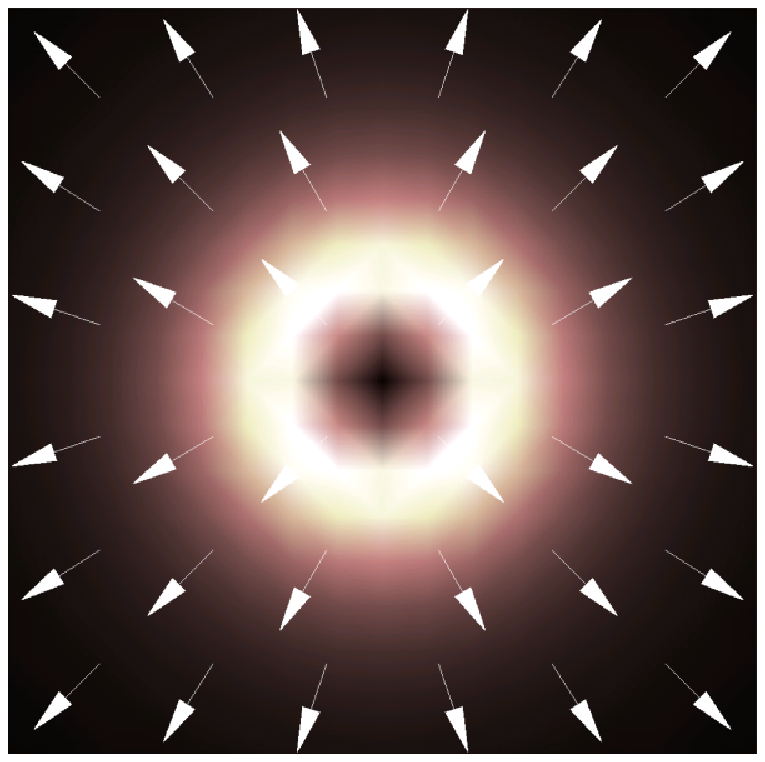}
\includegraphics[height=2.75in]{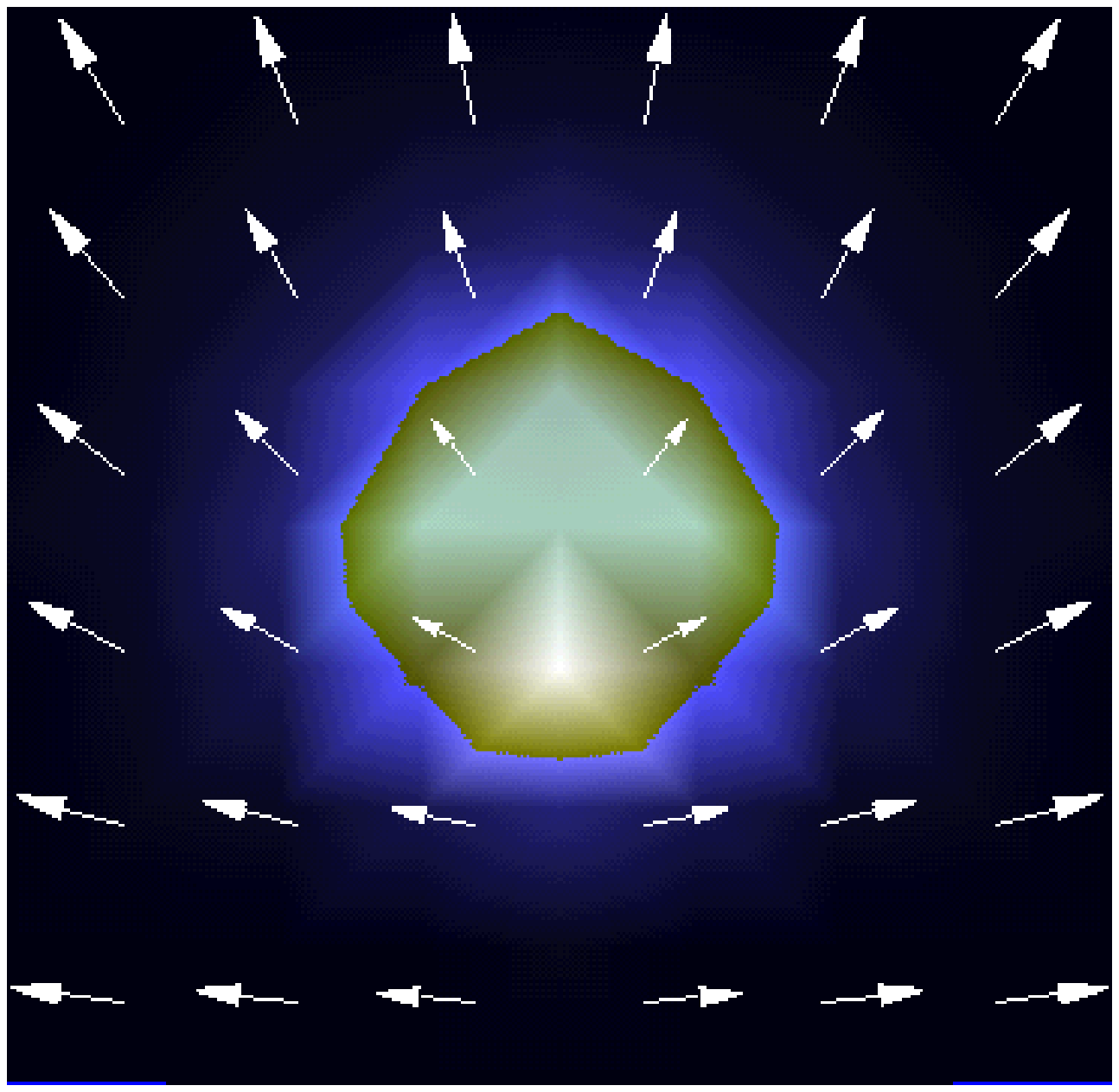} 
\end{center}
\caption{a) (left) $t = 9.2 \mu s$ (corresponding to the spherulites from Figures \ref{figtext}i-ii, actual data taken from the 25 layer simulation) the biaxial halo and uniaxial center of a single $+1$ disclination observed b) (right) $t = 28.7 \mu s$ (corresponding to the 50 layer simulation, Figure \ref{figtext}iii) the biaxial core of a $+\frac{1}{2}$ disclination observed after splitting \cite{Yan2002}; the surface corresponds to the degree of biaxiality (\ref{eqbiax}), arrows correspond to the uniaxial director (and should be considered headless), and the domain length scale is approximately $l= 4 nm$, or approximately one smectic layer, for both figures.  \label{figtextdefect}}
\end{figure}

The shape dynamic growth regime of this spherulite morphology has been found to have topologically similar dynamics compared to the growth of initially radial nematic spherulites \cite{Wincure2006,Wincure2007,Wincure2007a,Wincure2007b}.  In agreement similar nematic growth simulations \cite{Wincure2006,Wincure2007,Wincure2007a,Wincure2007b}, the initially imposed isotropic core is observed to undergo a self-selected transition to a $+1$ disclination, shown in Figure \ref{figtextdefect}a.  This initial morphology minimizes elastic long-range energy through the formation of a high-energy defect.  As the spherulite radius exceeds the smectic coherence length (\ref{eqn_smeclen}), long-range elastic gradients become energetically favorable compared to gradients in short-range order.  Subsequently, the $+1$ disclination splits into a pair of $+\frac{1}{2}$ disclinations \cite{Schopohl1987,Yan2002}.  This is a topologically equivalent texture that minimizes short-range energy, where the energy of a defect is proportional to its the square of its strength $s^2$ \cite{Rey2002}.  The smectic-A layer configurations corresponding with this splitting process is shown in Figure \ref{figtextzoom}.

The 25 layer simulation, with a more refined mesh, was used to both verify mesh independence of the 50 layer simulation (see Section \ref{sec_selfsim}) and obtain a more refined view of the smectic disclinations observed in this work.  In order to identify disclination defects, the degree of biaxiality (see Section \ref{subsec_elemdisl}) can be computed as follows \cite{Kaiser1992,Kralj1999}:
\begin{equation} \label{eqbiax}
\beta^2 = 1-6\frac{\left[\left(\bm{Q}\cdot\bm{Q}\right) : \bm{Q}\right]^2}{\left(\bm{Q} : \bm{Q}\right)^3}
\end{equation}
The core of the $+1$ disclination is shown in Figure \ref{figtextdefect}a as was determined from the 25 layer simulation.  The simulated structure of this disclination is both topologically correct and its biaxial halo/uniaxial center structure well-reproduced, compared to its nematic counterpart \cite{Hudson1993,Sonnet1995,Sigillo1998,Kralj1999,Andrienko2000}.  The spherulite length scale of the 25 layer simulation was not large enough to observe the splitting of the $+1$ disclination into a pair of $+\frac{1}{2}$, thus Figure \ref{figtextdefect}b shows the partially biaxial core structure of the $+\frac{1}{2}$ computed in the 50 layer simulation.  The core of this defect is also in agreement with its nematic counterpart, with a solely biaxial core \cite{Schopohl1987}.

\subsubsection{Self-similar growth} \label{sec_selfsim}

\begin{figure}[htp]
\begin{center}
\includegraphics[width=4in]{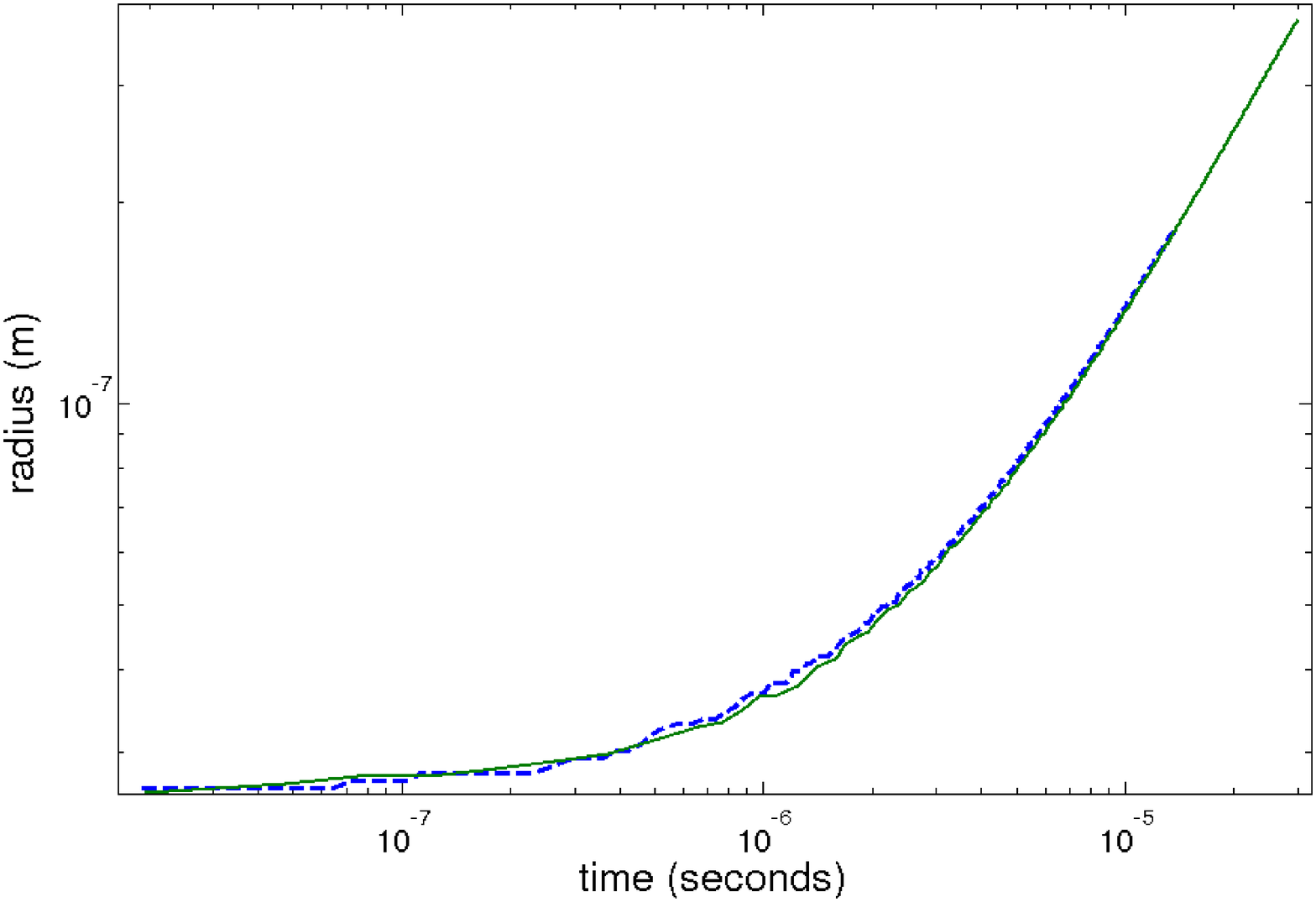}
\end{center}
\caption{Log-log plot of the spherulite radius versus time for the 25 layer (dotted line) and 50 layer (solid line) simulations where a power law fit ($r=t^n$) of the final data points (50 layer simulation) yields $n=0.95$. \label{figtextplaw}}
\end{figure}

The radius of the spherulite versus time was determined for both simulations.  Figure \ref{figtextplaw}a shows this data and a comparison of the results from the 50 layer and more refined 25 layer simulations show good agreement.  Deviation in the early stages of growth result from the more refined 25 layer simulation capturing the formation and core structure of the $+1$ disclination more accurately.  The convergence of the evolution of the two spherulite radii establishes the accuracy of the 50 layer simulation results beyond the radius of convergence.

Following the shape dynamic growth regime, the spherulite forms a scale-independent, or self-similar, shape \cite{Sethna2007}.  Spherulite growth obeys a power law $r=t^n$ in this growth regime, where $n$ is both experimentally observed \cite{Dierking2001,Dierking2003} and theoretically predicted \cite{Wincure2007,Huisman2007} to approach $n=1$ for phase-ordering processes quenched below the lower stability limit of the isotropic phase.  A power law fit of the final two points of the spherulite radius evolution (Figure \ref{figtextplaw}, 50 layer simulation) shows that the simulation results in this work correctly predict a self-similar growth regime, where $n=0.95$ is found.  Additionally, the results from Figure \ref{figtextplaw} show that the transition from the initial shape dynamics regime to the self-similar growth regime occurs at a relatively small length scale compared to the growth of initially radial textured nematic spherulites \cite{Wincure2007}.

\begin{figure}[htp]
\begin{center}
\includegraphics[width=4in]{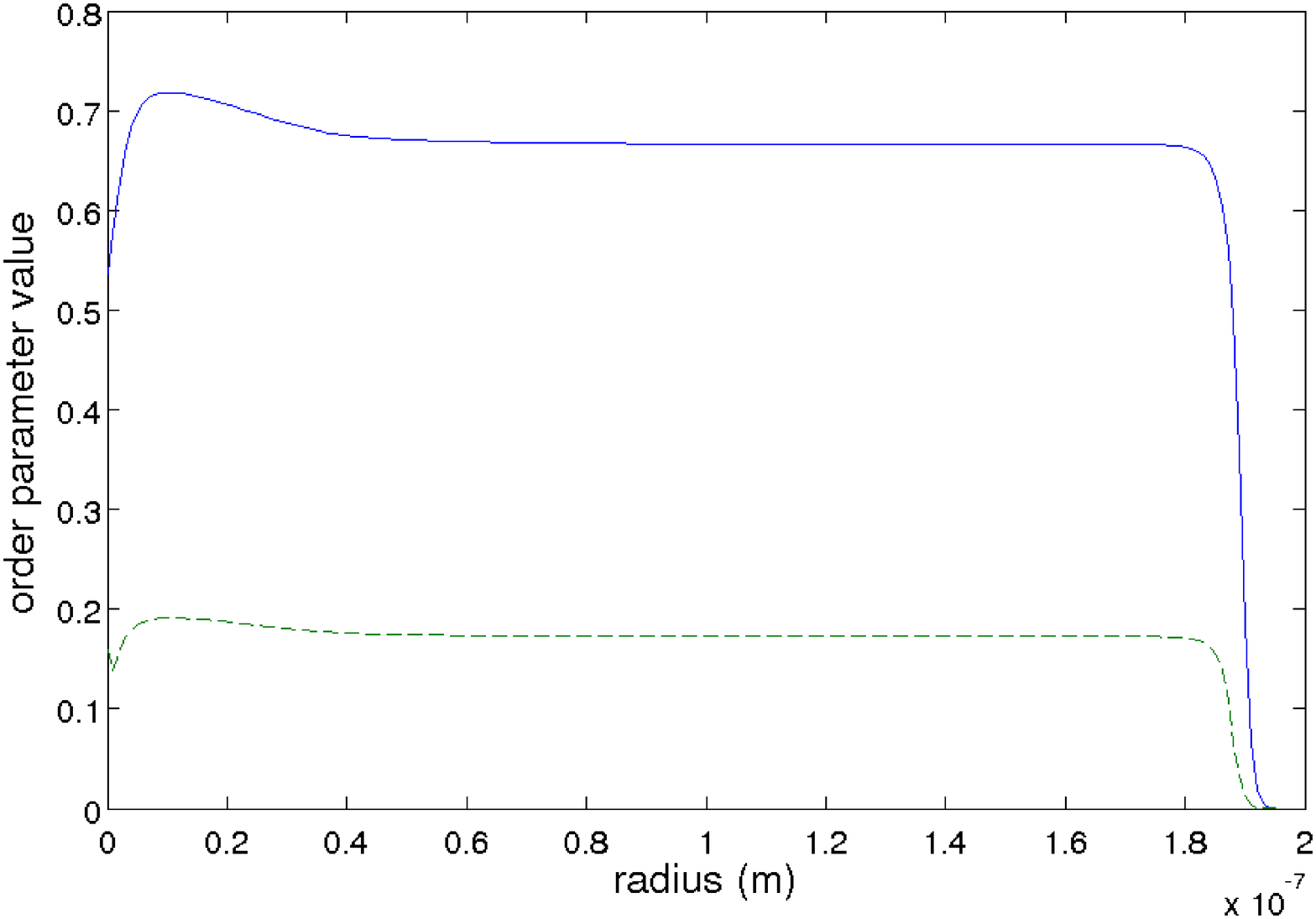}
\end{center}
\caption{$t = 28.7 \mu s$ (corresponding to the 50 layer simulation, Figure \ref{figtext}iii) Order parameter profiles of $S$ (solid line) and $\psi$ (dotted line); the symmetry axis perpendicular to that formed by the $+\frac{1}{2}$ disclination pair was used and bulk values determined in Section \ref{resdisc_phase} for $T=330.0K$ are $S=0.75$ and $\psi=0.20$. \label{figordprof}}
\end{figure}

Figure \ref{figordprof} shows the order parameter profiles during the beginning of the self-similar growth regime (Figure \ref{figtext}iii).  Good agreement is observed between the bulk values of the order parameters (determined in Section \ref{resdisc_phase}) and those found in the dynamic simulation.  The imposition of layer compression/expansion and curvature from the spherulite morphology and interfacial anchoring result in a decrease of the smectic-A order from that of the ideal homogeneous layer configuration.  Additionally, a local increase in smectic-A order is observed in the region immediately outside of the spherulite core where layer compression/expansion is minimized by the presence of the defect structure within the core (See Figure \ref{figtextzoom}iii).  This phenomenon is an example of the competition between gradients in long- and short-range order.  The existence of such high smectic-A ordering in the presence of substantial layer curvature is in agreement with both theoretical and experimental observations that layer bending is a low energy distortion in the smectic-A mesophase \cite{deGennes1995,Kleman2003}.

\section{Conclusions} \label{sec_conc}

A comprehensive approach to modeling the first-order isotropic/smectic-A phase transition was presented and applied to phase diagram computation and growth of an initially radial smectic-A spherulite in an isotropic phase.  Summarizing the results determined from this work are as follows:\begin{itemize}
\item An optimized method of phase diagram computation was presented and applied (Section \ref{sec_phasediag}), showing good agreement (Figure \ref{figphasediag}) with the experimental data used in parameter determination (Section \ref{subsec_params}).
\item Shape dynamics in the early spherulite growth period were found in agreement with the past theoretical work on a similar system, the first-order isotropic/nematic transition \cite{Yan2002,Sharma2003,Wincure2006,Wincure2007,Wincure2007a,Wincure2007b,Abukhdeir2008b}, where the initial imposed isotropic core forms a self-selected $+1$ disclination.  As the spherulite radius increases, this $+1$ disclination splits into two $+\frac{1}{2}$ disclinations, which is governed by a competition of long- and short-range ordering in the presence of interfacial anchoring \cite{Rey2002,Yan2002} (Figures \ref{figtext}-\ref{figtextdefect}).
\item A self-similar spherulite growth regime was observed following the initial shape dynamic growth regime where a power law fit of the spherulite radius was shown to approach $n=1$ (Figure \ref{figtextplaw}), in agreement with past experimental \cite{Dierking2001,Dierking2003} and theoretical \cite{Wincure2007,Huisman2007} studies of mesophase spherulite growth under a ``deep'' quench.
\item The general structure of both $+1$ and $+\frac{1}{2}$ smectic disclinations were found in agreement with studies of their nematic counterparts \cite{Schopohl1987,Hudson1993,Sonnet1995,Sigillo1998,Kralj1999,Andrienko2000}.  Note that due to computational constraints, the mesh density used in this simulation was adequate only to the extent of resolving the general structure of the defect core and not resolution with full detail.
\end{itemize}  The simulation results presented show that the use of a high-order phenomenological model and experimentally based model parameters results in a substantially more complete reproduction of the physical smectic-A system.  Though current computational resources restrict this initial work to two-dimensions, these results show a strong correlation to past experimental \cite{Schopohl1987,Hudson1993,Sonnet1995,Sigillo1998,Kralj1999,Andrienko2000,Dierking2001,Dierking2003} and theoretical observations \cite{Yan2002,Sharma2003,Wincure2006,Wincure2007,Wincure2007a,Wincure2007b,Huisman2007,Abukhdeir2008b}.  These promising results show that three-dimensional simulation of this model could be used to study the formation and dynamics of smectic-A spherulites at length and time scales currently inaccessible via experimental study.

\section*{Acknowledgments}
This work was supported, in part, by a grant from the Natural Science and Engineering Council of Canada.

\bibliographystyle{nature}
\bibliography{/home/nasser/nfs/references/references}

\end{document}